\documentclass[11pt]{article}
\usepackage{amssymb}
\usepackage{amsmath}
\usepackage{amscd}
\usepackage{latexsym}
\usepackage{bbm}

\catcode`@=11
\renewcommand{\section}{\@startsection{section}{1}{0pt}{\medskipamount}
{\medskipamount}{\large\bf}}
\numberwithin{equation}{section}
\catcode`@=12

\def\a{\alpha}

\def\de{\delta}
\def\eps{\epsilon}

\def\th{\theta}

\def\ka{\kappa}
\def\la{\lambda}

\def\p{\phi}

\newcommand{\C}{\mathbb C}
\newcommand{\R}{\mathbb R}

\newcommand{\NN}{{\mathbbm{N}}}
\newcommand{\unity}{\mathbbm{1}}
\newcommand{\Hcal}{{\mathcal H}}
\newcommand{\adag}{a^{\dagger}}

\def\im{\textrm{i}}
\def\ep{\textrm{e}}
\def\diff{\textrm{d}}

\def\Tr{\textrm{Tr}}
\def\tU{\textrm{U}}

\def\sfrac#1#2{{\textstyle\frac{#1}{#2}}}

\def\>{\rangle}
\def\<{\langle}
\def\+{\dagger}
\def\={\ =\ }
\def\zb{{\bar{z}}}
\def\and{\quad\textrm{and}\quad}
\def\und{\qquad\textrm{and}\qquad}

\begin{document}

\begin{titlepage}
\setcounter{page}{0}
\begin{flushright}
ITP--UH--16/13
\end{flushright}

\vskip 2.0cm

\begin{center}

{\LARGE\bf  Exact BPS bound for noncommutative baby Skyrmions}

\vspace{12mm}

{\LARGE Andrei Domrin,$\!{}^*$  
 \LARGE Olaf Lechtenfeld,$\!{}^+$ 
 \LARGE Rom\'an Linares,$\!{}^\times$ 
 \LARGE Marco Maceda${}^\times$
}
\\[8mm]
\noindent ${}^*${\em 
Department of Mathematics and Mechanics, Moscow State University\\
Leninskie gory, 119992, GSP-2, Moscow, Russia}\\
Email: \texttt{domrin@mi.ras.ru}
\\[8mm]
\noindent ${}^+${\em 
Institut f\"ur Theoretische Physik and
Riemann Center for Geometry and Physics\\
Leibniz Universit\"at Hannover,
Appelstra\ss{}e 2, 30167 Hannover, Germany }\\
Email: \texttt{lechtenf@itp.uni-hannover.de}
\\[8mm]
\noindent ${}^\times${\em 
Departamento de F\'isica, Universidad Aut\'onoma Metropolitana Iztapalapa \\
San Rafael Atlixco 186, C.P.~09340, M\'exico D.F., M\'exico }\\
Emails: \texttt{\{ lirr, mmac \} @xanum.uam.mx}

\vspace{12mm}

\begin{abstract}

\noindent
The noncommutative baby Skyrme model is a Moyal deformation of the
two-dimensional sigma model plus a Skyrme term, with a group-valued 
or Grassmannian target. Exact abelian solitonic solutions have been 
identified analytically in this model, with a singular commutative
limit. Inside any given Grassmannian, we establish a BPS bound for 
the energy functional, which is saturated by these baby Skyrmions.
This asserts their stability for unit charge, as we also test in
second-order perturbation theory.
\end{abstract}

\end{center}
\end{titlepage}

\section{Introduction and summary}

\noindent
The baby Skyrme model is a useful laboratory for studying soliton physics.
It is the $2{+}1$ dimensional analog of the usual Skyrme model~\cite{S}, which
describes the low-energy chiral dynamics of quantum chromodynamics~\cite{ANW}.
This model has direct applications in condensed matter physics~\cite{Mac}, 
where baby Skyrmions give an effective description in quantum Hall systems. 
The action of this model consists of three terms: a kinetic sigma-model term
(scale invariant), the (four-derivative) Skyrme term (breaking scale invariance)
and a potential (or mass) term (stabilizing the size of solutions).
All three terms are needed to prevent the collapse of topological configurations 
which yield to Skyrmion solutions. These stable baby Skyrmions can be determined
numerically~\cite{Zak}. Their mass is strictly larger than the Bogomol'nyi bound
given by the topological charge (Skyrmion number), and the two-Skyrmion
configuration becomes stable showing the existence of bound states~\cite{Zak}.

A noncommutative deformation~(for reviews see~\cite{nc})
serves as a substitute for the potential term,
because it introduces a new length scale into the theory, which also
stabilizes solitons against collapse or spreading.
Moreover, Moyal-deformed field theories have a much richer soliton spectrum
than their commutative counterparts 
(see, e.g.,~\cite{lepo01,sendai} and references therein).
Indeed, the noncommutativity gives rise to a new class of baby Skyrmions,
as was shown in~\cite{iole}.
Furthermore, the noncommutative deformation may be of help in semi-classically
quantizing the (perturbatively non-renormalizable) baby Skyrme model, since it 
introduces a regulating parameter.
The two above applications of noncommutativity are our main motivation
for Moyal-deforming the baby Skyrme model.

In a previous paper~\cite{iole} by one of the authors on this subject, 
the Moyal-deformed baby Skyrme model was introduced~\footnote{
See also \cite{mieck} for different aspects of Moyal-deforming a Skyrme model.}
for group-valued or Grassmannian target spaces and without a potential term.
In the abelian case, a class of exact analytic solitonic solutions was discovered,
which are stable against scaling due to the noncommutativity but have no analogues 
in the commutative theory. This surprising feat succeeded because certain BPS 
configurations of the Moyal-deformed ordinary sigma model extremize the Skyrme part 
of the energy as well. The static energy of these noncommutative baby Skyrmions
and their repulsive potential at large distances was computed~\cite{iole}. 
However, their stability could not be ascertained, because a BPS bound for
the full baby Skyrme model (in a given Grassmannian) was not available.~\footnote{
For the pure sigma model, the energy is of course bounded by the topological
charge~\cite{dolepe}. The Skyrme term together with a potential also enjoys
a BPS bound which, however, becomes trivial for zero potential~\cite{gipa,adam}.}

In the present Letter, we fill this gap. 
After reviewing the salient features of the noncommutative baby Skyrme model and 
its known solutions, we prove the expected BPS bound for the Skyrme term in the 
energy functional. The special case of unit topological charge is established 
independently by mapping it to the quantum mechanical uncertainty relation. 
Finally, we develop the second-order perturbation of the energy functional 
around a classical solution and apply it to the charge-one baby Skyrmion, 
affirming our previous results.

\bigskip

\section{The noncommutative abelian baby Skyrme model}
\noindent
The Moyal-deformed baby Skyrme model was first introduced in~\cite{iole}.
Its abelian version describes maps~$g$ from a time interval $I\ni t$ into the unitaries
$\tU(\Hcal)$ of a Hilbert space~$\Hcal$ %(a Heisenberg-algebra representation module)
or into a Grassmannian subspace 
\begin{equation}
\textrm{Gr}_k\equiv\textrm{Gr}(P)\=\frac{\tU(\Hcal)}{\tU(\textrm{im}P)\times\tU(\textrm{ker}P)}
\end{equation}
for a hermitian projector~$P$ of finite rank~$k$.
In other words, the field variable $g(t)$ is a unitary operator-valued function of time.
Inside the Grassmannian Gr${}_k\subset\tU(\Hcal)$, it satisfies the constraint
\begin{equation} \label{Gr}
g^2=\unity \qquad\Leftrightarrow\qquad g^\+=g \qquad\Leftrightarrow\qquad
g=\unity-2P \quad\textrm{with}\quad P^\+=P=P^2\ ,
\end{equation}
defining a hermitian projector~$P(t)$ of rank~$k$ as an alternative field variable.
The Hilbert space~$\Hcal$ carries a representation of the Heisenberg algebra,
\begin{equation} \label{heisenberg}
[\,a\,,\,\adag\,] \= \unity\ ,
\end{equation}
which acts on the orthonormal basis states
\begin{equation}
|m\>\=\sfrac{1}{\sqrt{m!}}\,(\adag)^m\,|0\>
\qquad\text{for}\quad m\in\NN_0 \quad\text{and}\quad a|0\>=0
\end{equation}
in the following way,
\begin{equation}
a\,|m\> \= \sqrt{m}\,|m{-}1\> \ ,\qquad
\adag\,|m\> \= \sqrt{m{+}1}\,|m{+}1\> \ ,\qquad
N\,|m\> \ := \adag a\,|m\> \= m\,|m\>\ .
\end{equation}

With the help of the auxiliary gauge potentials
\begin{equation}
A_t=g^\+\dot{g} \und 
A_z=g^\+[\adag,g] \qquad\textrm{as well as}\qquad A_\zb=g^\+[a\,,g]=(A_z)^\+\ ,
\end{equation}
the model is defined by its action functional,
\begin{equation}
S\=-2\pi\!\int\!\!\diff t\ \Tr_{\Hcal} \Bigl\{
\sfrac{\th}{2}A_t^2\ +\ A_z A_\zb\ -\ 
\ka^2[A_t,A_z][A_t,A_\zb]\ +\ \sfrac{\ka^2}{2\th}[A_z,A_\zb]^2 \Bigr\}\ ,
\end{equation}
which depends on two parameters: the noncommutativity scale~$\th\in\R_+$ of the 
dimension of length$^2$ and a coupling parameter~$\ka$ of the dimension of length.
Note that no potential term is needed, because the presence of the scale~$\th$
stabilizes the solitonic solutions.
In the limit $\th{\to}0$, which includes scaling away the central charge
of the Heisenberg algebra~(\ref{heisenberg}), one recovers the commutative U(1) 
baby Skyrme model on $\R^{1,2}$, which is a free theory because all commutators vanish.
Sending the Skyrme coupling $\ka{\to}0$ also removes the quartic terms, leaving us
with the Moyal-deformed abelian sigma model.
The latter has been investigated intensively and features static BPS solitons
(see, e.g.~\cite{dolepe,klalepe}).

In this paper we are concerned with static solutions to the equation of motion,
$\dot{g}=0$. These extremize the energy functional
\begin{equation}
\begin{aligned}
E &\= 2\pi\,\Tr_{\Hcal} \Bigl\{ A_z A_\zb\ +\ \sfrac{\ka^2}{2\th}[A_z,A_\zb]^2 \Bigr\} 
\ =:\ E_0 + \sfrac{\ka^2}{\th}E_1 \\[6pt]
&\= 8\pi\,\Tr_{\Hcal} \Bigl\{ Q\,\adag P\,a+Q\,a\,P\,\adag \Bigr\} \\
&\ +\ 32\pi\sfrac{\ka^2}{\th}\,\Tr_{\Hcal} \Bigl\{
P\,a\,Q\,\adag P\,a\,Q\,\adag + P\,\adag Q\,a\,P\,\adag Q\,a - 
P\,a\,Q\,\adag P\,\adag Q\,a - P\,\adag Q\,\adag P\,a\,Q\,a \Bigr\}
\end{aligned}
\end{equation}
which, for later convenience, we have expressed in terms of the projectors
\begin{equation}
P \and Q=\unity{-}P \qquad\textrm{via}\qquad
A_z=-2(Q\,\adag P+P\,\adag Q) \und A_\zb=-2(Q\,a\,P+P\,a\,Q)\ .
\end{equation}
The energy depends only on the dimensionless combination~$\sfrac{\ka^2}{\th}$.

It was shown in~\cite{iole} that the diagonal projectors
\begin{equation} \label{Pk}
P^{(k)} \ :=\ \sum_{n=0}^{k-1}\,|n\>\<n|
\end{equation}
and their translates 
\begin{equation}
P^{(k|\a)}\ :=\ \ep^{\a\adag-\bar\a a}\,P^{(k)}\,\ep^{-\a\adag+\bar\a a}
\qquad\textrm{for}\quad \a\in\C \quad\textrm{and}\quad k\in\NN
\end{equation}
extremize both $E_0$ and $E_1$.\footnote{
Actually, one can show that {\sl any\/} diagonal projector solves the baby Skyrme
equation of motion.}
The Moyal deformation is essential for this property; 
in the commutative (nonabelian) case, sigma-model BPS solitons can never 
obey the baby Skyrme equation of motion.
The projector $P^{(k|\a)}$ can be interpreted
(via the Moyal-Weyl map) as a localized rank-$k$ baby Skyrmion, formed by
$k$~rank-one baby Skyrmions sitting on top of each other.
These configurations form a complex one-parameter
subfamily inside the complex $k$-parameter family of BPS~projectors for the 
noncommutative abelian sigma model (at $\ka{=}0$), where they saturate the bound
\begin{equation} \label{E0bound}
E_0 \= 8\pi\,\Tr_{\Hcal} \bigl\{Q\adag\!Pa+QaP\adag\!\bigr\}
\= 8\pi\,\Tr_{\Hcal} \bigl\{ P + 2\,QaP\adag\!\bigr\} 
\= 8\pi k\,+\,16\pi\,\Tr_{\Hcal} |Q\,aP|^2 \ \ge\ 8\pi k\ .
\end{equation}
No such bound was known for $E_1$, 
but the full energy of $P^{(k|\a)}$ was easily computed~\cite{iole},
\begin{equation}
E[P^{(k|\a)}]\= 8\pi\,\bigl(k + 4\sfrac{\ka^2}{\th}k^2\bigr)\ ,
\end{equation}
and is independent of~$\alpha$. The ensueing inequality
\begin{equation}
E[P^{(k|\a)}] \ \ge\ E[P^{(1|\a_1)}]+E[P^{(1|\a_2)}]+\ldots+E[P^{(1|\a_k)}]
\=k\,E[P^{(1)}]
\end{equation}
signals an instability of the localized rank-$k$ baby Skyrmion against decay 
into its constituents, a collection of $k$ well-separated rank-one baby Skyrmions.
Indeed, a repulsive force between two rank-one baby Skyrmions was found in~\cite{iole}.
General multi-center BPS~solitons of the $\ka{=}0$ sigma model do not solve the 
baby Skyrme equation of motion, but approach a classical solution for 
near-infinite mutual separation. This observation suggests a BPS bound also for
the Skyrme term,
\begin{equation} \label{E1bound}
E_1\ \ge\ 32\pi k\ .
\end{equation}
We will establish this bound in the following section.

\bigskip

\section{BPS bound for the Skyrme term}
\noindent
It is well known that, inside the full group of $\tU(\Hcal)$, one can connect any
Grassmannian solution to the vacuum via
\begin{equation}
g(s) \= \ep^{\im(\pi-s)P} \= \unity\ -\ (1{+}\ep^{-\im s})P
\qquad\textrm{with}\quad P^\+=P=P^2 \quad\textrm{and}\quad s\in[0,\pi]\ ,
\end{equation}
which monotonically decreases the energy from that of $g(0)=\unity{-}2P$
to the zero value of the vacuum $g(\pi)=\unity$~\cite{iole}.
Therefore, noncommutative baby Skyrmions can be stable only in the Grassmannian models.
Moreover, in Gr$_k$, only configurations of $k$~well-separated rank-one baby Skyrmions
have a chance to be stable, as we argued above. 

To prove this assertion, we rewrite the energy functional as
\begin{equation}
E\=8\pi\,\Tr_{\Hcal} \Bigl\{ |F|^2 + |G|^2 \ +\ 
2\sfrac{\ka^2}{\th} \bigl(F\,F^\+ - G^\+ G\bigr)^2 +
2\sfrac{\ka^2}{\th} \bigl(F^\+ F - G\,G^\+\bigr)^2 \Bigr\}
\end{equation}
with the abbreviations
\begin{equation}
F = P\,a\,Q \and G = Q\,a\,P \qquad\Rightarrow\qquad
F^\+ = Q\,\adag P \and G^\+ = P\,\adag Q\ .
\end{equation}
The positivity of this expression is obvious, but improving the lower bound requires 
using the Heisenberg algebra~(\ref{heisenberg}) and the topological charge formula
\begin{equation}
\Tr_{\Hcal}\bigl\{F^\+ F - G\,G^\+\bigr\} \=
\Tr_{\Hcal}\bigl\{F\,F^\+ - G^\+ G\bigr\} \=
\Tr_{\Hcal}\bigl\{P\,a\,\adag-P\,\adag a\bigr\}\= k\ .
\end{equation}
Note that all four operators
\begin{equation}
F\,F^\+\ ,\quad F^\+ F\ ,\quad G\,G^\+ \and G^\+ G
\end{equation}
are hermitian and non-negative definite with a rank at most equal to~$k$.
Therefore, the spectral theorem guarantees that both differences
$F\,F^\+ - G^\+ G$ and $F^\+ F - G\,G^\+$ have, 
in appropriate orthonormal bases, the form
\begin{equation}
\textrm{diag}\bigl(\la_1,\la_2,\ldots,\la_\ell,
-\mu_1,-\mu_2,\ldots,-\mu_m,0,0,\ldots\bigr)
\qquad\textrm{with}\qquad 
\sum_{i=1}^\ell \la_i-\sum_{j=1}^m \mu_j=k\ ,
\end{equation}
where $\la_i,\mu_j>0$ and $\ell+m\le 2k$. It may happen that $m=0$ (no negative 
eigenvalues), but always $\ell\ge 1$ (since the trace is positive). We claim that
$\ell\le k$. Indeed, in the first case,
\begin{equation}
\textrm{im}\bigl(F\,F^\+ - G^\+ G\bigr)\,\subseteq\,\textrm{im}\,P 
\qquad\Rightarrow\qquad   \textrm{rk}\bigl(F\,F^\+ - G^\+ G\bigr)\le k\ , 
\end{equation}
so that the stronger condition $\ell{+}m\le k$ holds.
In the second case, 
\begin{equation}
\textrm{im}\bigl(F^\+ F-G\,G^\+\bigr)\,\subseteq\,\textrm{im}\,Q\ ,
\end{equation}
and $F^\+ F-G\,G^\+$ is obviously non-positive definite on~$\textrm{ker}\,F$.
But $\textrm{ker}\,F$ is the orthogonal complement to $\textrm{im}\,F^\+$ and, 
therefore, has codimension at most equal to~$k$.
In case $\ell>k$, it would have a non-zero intersection with the $\ell$-dimensional
linear span of all eigenvectors of $F^\+ F-G\,G^\+$ corresponding to the positive
eigenvalues $\la_1,\la_2,\ldots,\la_\ell$. The resulting contradiction shows
that $\ell\le k$ in the second case as well.

To prove our inequality~(\ref{E1bound}), we have to estimate the trace of the square
of the two difference operators, which in each case is given by
\begin{equation}
\sum_i\la_i^2+\sum_j\mu_j^2 \qquad\textrm{subject to}\qquad
\sum_i\la_i-\sum_j\mu_j=k \und \la_i,\mu_j>0\ .
\end{equation}
Implementing the first subsidiary condition via Lagrange multipliers in the variational problem,
one sees that the existence of extrema is in contradiction with the positivity of the~$\mu_j$.
Therefore, a minimum is attained for any $\ell\le k$ but only for $m{=}0$ 
(no negative eigenvalues) and at
\begin{equation}
\la_1=\la_2=\cdots=\la_\ell=\sfrac{k}{\ell} \qquad\Rightarrow\qquad
\sum_i\la_i^2\ \ge\ \ell\bigl(\sfrac{k}{\ell}\bigr)^2\ \ge\ k\ .
\end{equation}
This bound is saturated only for $\ell{=}k$, 
i.e.~when there are precisely $k$~eigenvalues of magnitude one.
We have thus shown that
\begin{equation}
\Tr_{\Hcal}\Bigl\{\bigl(F\,F^\+ - G^\+ G\bigr)^2\Bigr\}\ \ge\ k \und
\Tr_{\Hcal}\Bigl\{\bigl(F^\+ F - G\,G^\+\bigr)^2\Bigr\}\ \ge\ k\ ,
\end{equation}
and (\ref{E1bound}) follows. 
The complete bound in Gr$_k$ then reads
\begin{equation} \label{fullbound}
E\ \ge\ 8\pi\,k\,\bigl(1+4\sfrac{\ka^2}{\th}\bigr)\ .
\end{equation}

This confirms the exclusive stability of the noncommutative
abelian rank-one baby Skyrmion and widely separated collections of them,
\begin{equation}
\begin{aligned}
P^{(k|\a_1,\a_2,\ldots,\a_k)} &\= \sum_{i,j=1}^k
|\a_i\> \,\bigl( \<\a_.|\a_.\> \bigr)^{-1}_{ij} \<\a_j|
\qquad\textrm{for}\qquad \a_i\in\C \und |\a_i{-}\a_j|\to\infty \\
&\ \approx\ \sum_i \ep^{-|\a_i|^2} |\a_i\> \<\a_i|\ ,
\end{aligned}
\end{equation}
employing $k$ coherent states defined by \
$|\a_i\>=\ep^{\a_i a^\+}\,|0\>$ \
and the matrix of their overlaps~$\<\a_i|\a_j\>$.
These are the only configurations saturating the BPS bound~(\ref{fullbound}).

The rank-one case Gr$_1$ is critical, so let us give it a different look.
Any rank-one hermitian projector is determined by a state vector~$|\psi\>\in\Hcal$,
\begin{equation}
P \= |\psi\>\<\psi| \qquad\textrm{with}\quad \<\psi|\psi\>=1\ .
\end{equation}
After some algebra, the energy functional in Gr$_1$ takes the following form,
\begin{equation} \label{Erk1}
\begin{aligned}
E &\= 8\pi\bigl\{ \<a\,\adag\> + \<\adag a\,\> \bigr\} \ +\
32\pi\sfrac{\ka^2}{\th}\bigl\{ 1+\<a\,\adag\>\<\adag a\,\>-\<a\,a\,\>\<a^\+a^\+\> \bigr\} \\
&\= 8\pi\bigl\{ \<x^2\> + \<p^2\> \bigr\} \ +\
32\pi\sfrac{\ka^2}{\th}\bigl\{ \sfrac34 + \<x^2\>\<p^2\> - \sfrac14\<xp{+}px\> \bigr\}\ ,
\end{aligned}
\end{equation}
with the {\sl connected\/} expectation values
\begin{equation}
\<Y\>=\<\psi|Y|\psi\> \und \<YZ\>=\<\psi|YZ|\psi\>-\<\psi|Y|\psi\>\<\psi|Z|\psi\>\ .
\end{equation}
In the second line of~(\ref{Erk1}), we expressed the raising and lowering operators
through the hermitian combinations $x$ and~$p$ (quantum mechanical position and momentum),
\begin{equation}
a \= \sfrac1{\sqrt{2}}(x+\im p) \und \adag \= \sfrac1{\sqrt{2}}(x-\im p)
\qquad\Rightarrow\qquad [x,p]=\im\unity\ .
\end{equation}
The Robertson uncertainty relation~\cite{rob} of elementary quantum mechanics tells us that
\begin{equation}
\<x^2\>\<p^2\>\ \ge\ \bigl|\sfrac1{2\im}\< [x,p] \>\bigr|^2 \= \sfrac14 
\qquad\Rightarrow\qquad \<x^2\> + \<p^2\> \ \ge\ 1\ ,
\end{equation}
which recovers the familiar bound~(\ref{E0bound}) for~$E_0$. To estimate~$E_1$, 
we need the (stronger) Schr\"odinger uncertainty relation~\cite{sch},\footnote{
We are grateful to Reinhard F.~Werner for the hint.}
\begin{equation}
\<x^2\>\<p^2\>\ \ge\ 
\bigl|\sfrac12\<\{x,p\}\>\bigr|^2\ +\ \bigl|\sfrac1{2\im}\< [x,p] \>\bigr|^2
\qquad\Rightarrow\qquad \<x^2\>\<p^2\>-\sfrac14\<xp{+}px\>\ \ge\ \sfrac14\ ,
\end{equation}
which bounds the second curly bracket on each line of~(\ref{Erk1}) by~1
and thus yields $E_1\ge 32\sfrac{\ka^2}{\th}$, as anticipated.
Mathematically, it is nothing but the Cauchy-Schwarz inequality at work.

\bigskip 

\section{Second-order perturbation around baby Skyrmions}
\noindent
It is instructive to study the energy functional in the neighborhood of a classical 
solution~$g$. In order to remain inside the Grassmannian, where $g^\+=g=\unity{-}2P$,
we set up a multiplicative perturbation expansion,
\begin{equation}
g(\eps) \= g\,\ep^\p \qquad\textrm{with}\qquad \p^\+=-\p\ ,\quad \{\p,g\}=0
\und \p=O(\eps)\ ,
\end{equation}
which is `odd' with respect to $P$ in the sense that
\begin{equation}
P\,\p =\p\,Q \and \p\,P=Q\,\p \qquad\Leftrightarrow\qquad \p = P\,\p +\p\,P\ .
\end{equation}
To second order in the perturbation, we compute
\begin{equation}
P(\eps)\=P\ -\ \sfrac12(\unity{-}2P)\bigl(\p+\sfrac12\p^2+O(\eps^3)\bigr)
\=P\ +\ \sfrac12[P,\p]\ +\ \sfrac18\bigl[[P,\p],\p\bigr]\ +\ O(\eps^3)
\end{equation}
and introduce the abbreviations
\begin{equation}
A=A_z=g\,[\adag,g]\ ,\quad \bar{A}=A_\zb=g\,[a\,,g]\ ,\quad
B=[\adag{+}A\,,\p]\ ,\quad \bar{B}=[a\,{+}\bar{A}\,,\p]\ .
\end{equation}
The equation of motion takes the form
\begin{equation} \label{eom}
[a\,,C]+[\adag,\bar{C}]\=0 \qquad\textrm{with}\qquad
C=A-\sfrac{\ka^2}{\th}\bigl[A\,,[A\,,\bar{A}]\bigr] \and
\bar{C}=\bar{A}-\sfrac{\ka^2}{\th}\bigl[\bar{A}\,,[\bar{A}\,,A]\bigr]\ .
\end{equation}

After a straightforward but lengthy calculation, the energy functional
inside Gr$_k$, expanded to second order in~$\eps$ around a classical
projector~$P$ subject to~(\ref{eom}), can be simplified to
\begin{equation}
\begin{aligned}
E[P(\eps)]\=E[P]\ +\ \pi\,&\Tr_{\Hcal} 
\bigl\{ 2\,B\bar{B}-[C,\p]\bar{B}-[\bar{C},\p]B\bigr\} \\
+\ 2\pi\sfrac{\ka^2}{\th}\,&\Tr_{\Hcal}\bigl\{ 
2B\bar{A}A\bar{B}+2\bar{B}A\bar{A}B-B\bar{A}\bar{A}B-\bar{B}AA\bar{B}-
BA\bar{A}\bar{B}-\bar{B}\bar{A}AB \\
&\qquad +B\bar{A}B\bar{A}+\bar{B}A\bar{B}A- B\bar{A}\bar{B}A-BA\bar{B}\bar{A} 
\bigr\} \ +\ O(\eps^3)\ .
\end{aligned}
\end{equation}
Note that $B$ and $\bar{B}$ contain~$\p$ and are thus of~$O(\eps)$,
and there is a hidden $\ka$~dependence in $C$ and~$\bar{C}$.

Let us evaluate this expression for the unique (up to translation) rank-one 
baby Skyrmion,
\begin{equation}
P^{(1)}\=|0\>\<0| \qquad\Rightarrow\qquad
A=-2\;|1\>\<0| \und C=\bigl(1{+}8\sfrac{\ka^2}{\th}\bigr)\,A\ ,
\end{equation}
and the most general perturbation inside~Gr$_1$,
\begin{equation}
\p\= \sum_{n=1}^{\infty}\Bigl\{ \p_n\,|0\>\<n|-\p^*_n\,|n\>\<0| \Bigr\}
\qquad\textrm{with}\quad \p_n\in\C\ .
\end{equation}
One finds that
\begin{equation}
B\=-\sum_{n=1}^{\infty}\Bigl\{
\p_n\,|1\>\<n|+\sqrt{n}\,\p_n\,|0\>\<n{-}1|-2\de_{n1}\p_1\,|0\>\<0|
+\sqrt{n{+}1}\,\p^*_n\,|n{+}1\>\<0| \Bigr\} 
\end{equation}
\begin{equation}
\!\:\textrm{and} \qquad
[C,\p]\=-2\bigl(1{+}8\sfrac{\ka^2}{\th}\bigr)\sum_{n=1}^{\infty}\Bigl\{
\p_n\,|1\>\<n|-\de_{n1}\p_1\,|0\>\<0| \Bigr\} 
\qquad\qquad\qquad\qquad{}
\end{equation}
and finally
\begin{equation} \label{Epert}
E[P^{(1)}(\eps)]\= 8\pi\bigl(1{+}4\sfrac{\ka^2}{\th}\bigr)\ +\ 8\pi\,|\p_2|^2
\ +\ 4\pi\bigl(1{+}2\sfrac{\ka^2}{\th}\bigr)\sum_{n=3}^{\infty}n\,|\p_n|^2
\ +\ O(\eps^3)\ .
\end{equation}
A $\p_1$ perturbation corresponds to the translational mode and does not cost
any energy. The Skyrme term does not see the $\p_2$ perturbation either.
Clearly, the bound~(\ref{fullbound}) for $k{=}1$ is respected.

One can go beyond perturbation theory by probing all basis directions in Gr$_1$
exactly,\footnote{
We suppress the possibility of adding relative phases in $|\psi_n(\eps)\>$ as well.}
\begin{equation} \label{Pn}
P^{(1)}_n(\eps)\=|\psi_n(\eps)\>\<\psi_n(\eps)| \qquad\textrm{with}\qquad
|\psi_n(\eps)\> \= \cos\eps\,|0\>+\sin\eps\,|n\> \und \eps\in[0,2\pi]\ .
\end{equation}
Inserting these projector families into~(\ref{Erk1}), we arrive at
\begin{equation}
E[P^{(1)}_n(\eps)]\= \begin{cases}
8\pi(1{+}2\sin^4\!\eps) + 32\pi\sfrac{\ka^2}{\th}(1{+}2\sin^6\!\eps)
& \quad \textrm{for} \quad n=1 \\
8\pi(1{+}4\sin^2\!\eps) + 32\pi\sfrac{\ka^2}{\th}(1{+}6\sin^4\!\eps)
& \quad \textrm{for} \quad n=2 \\
8\pi(1{+}2n\sin^2\!\eps) + 32\pi\sfrac{\ka^2}{\th}(1{+}n\sin^2\!\eps{+}n^2\sin^4\!\eps) 
& \quad \textrm{for} \quad n\ge3 
\end{cases}\ .
\end{equation}
To order~$\eps^2$, this precisely reproduces the coefficients of~$|\p_n|^2$ in~(\ref{Epert})
after matching $|\p_n|^2=4\eps^2$.
Again, it is apparent that only $P^{(1)}_n(0)=P^{(1)}$ is stable. Beyond $O(\eps^2)$, 
the flat valley traced by \ $P^{(1|\a)}=\ep^{-|\a|^2}|\a\>\<\a|$ \ deviates from the
curves defined in~(\ref{Pn}). 

We close with a list of open problems. It would be interesting to 
work out the scattering of two rank-one baby Skyrmions in the Moyal plane.
It is also an open question whether there exist abelian noncommutative 
baby Skyrmions not based on diagonal projectors.
Another promising task is to deform the {\it full\/} Skyrme model (on $\R^{1,3}$) 
and to construct noncommutative Skyrmions from noncommutative instantons~\cite{inst}.

%\medskip

\noindent
{\bf Acknowledgements}\\
\noindent
We are thankful for hospitality by UAM-Iztapalapa (O.L.) and by Leibniz University (R.L.~and M.M.).
Useful discussions with Mohab Abou Zeid and Reinhard F.~Werner are gratefully acknowledged.
This work is partially supported by DFG--CONACyT grants B330/285/11 and B330/418/11 as well as
by a DFG--RFFI collaboration grant LE 838/12-1. In addition, A.D.~was supported by the Russian 
Foundation for Basic Research under the grants 13-01-00622 and 13-01-12417.

%\medskip

\end{document}